\documentclass{article}
\usepackage{authblk}
\usepackage{amsmath}
\usepackage{amsfonts}
\usepackage{amssymb}
\usepackage{geometry}
\usepackage{scalefnt}
\usepackage{xcolor}
\usepackage[colorlinks=true,linkcolor=red,citecolor=blue,urlcolor=cyan]{hyperref}
\usepackage{graphicx,epstopdf}
\usepackage{subfigure}
\usepackage{float}
\usepackage{fullpage}
\numberwithin{equation}{section}
\providecommand{\U}[1]{\protect\rule{.1in}{.1in}}

\begin{document}

	\title{Nonextensive Effect on the Lump Soliton Structures in Dusty Plasma}
	\author[1] {Prasanta Chatterjee}
    \author[2*] {Uday Narayan Ghosh}
	\author[2] {Snehalata Nasipuri}
	\author[3]  {M.Ruhul Amin}
	\affil[1,2] {Department of Mathematics, Siksha Bhavana, Visva- Bharati, Santiniketan, Birbhum, West Bengal-731235, India}
	\affil[2*] {Department of Mathematics, KKM College, A Constituent unit of Munger, Bihar-811307, India}
	\affil[3] {Department of Mathematical and Physical Sciences, East West University, Aftabnagar, Dhaka-1212}
	\date{}
	\maketitle
\begin{abstract}
In this paper, we use a very prominent technique, Hirota Bilinear Method (HBM) to survey the lump structures of the Kadomtsev-Petviashvili (KP) equation in the frame of a collisionless magnetized plasma system composed of dust grains, ions, and nonextensive electrons. Nonlinearity has worldwide applications, and soliton theory is a powerful appliance to illustrate its qualitative behaviors. So, lump solitons are very significant and also interesting. We have observed that lump structures differ due to the correlated parameters of the plasma system. It has also been found that the nonextensive parameter crucially changes the lump features.
	\end{abstract}
    \section{Introduction}
    In advanced research in plasma physics, dusty plasma is a weighty and captivating research area. It is located in a major part of the space plasma and solar system. Due to its majority in space plasma  \cite{1}-\cite{5} and the solar system, it holds an enlarged portion of plasma research. It covers a broad area of space plasma and the solar system. So it plays a crucial role in the plasma physics research field. Dusty plasma becomes a hot research topic for its enormous exposition and some remarkable properties \cite{dp1}-\cite{dp6}. Dust grains are highly charged in the ionized medium. Dust grains are highly inactive and charge-centralized, so the united characteristics of the electromagnetic force field, which can  authorize the circulation of ion-acoustic solitary waves (IASWs), can be reformed by the charged dust grains. Subsequently, the dusty plasma produced some low-frequency collective states, which are remarkably different from the typical electron-ion plasmas. Experimental and theoretical examinations \cite{d10}-\cite{d14} confess the creation of low-frequency and highly low-frequency acoustic states, which have practical uses in space plasma along with laboratory plasma. \\

The interaction of materiality and recovering force generates various states. While the electron-ion pressure originates the recovering force, the materiality is raised by the excessive dust grains. It was first theoretically introduced by Rao et al. \cite{d10}  in unmagnetized dusty plasma that an ion-acoustic wave (IAW) is in such a variety of states. Dust ion-acoustic solitary waves (DIASWs) were predicted by Shukla and Silin \cite{di13} and Barkan et al. \cite{ka14} and Melrino et al. \cite{ka16} experimentally observed them in the laboratory. The dust acoustic shock waves \cite{pop1} were produced by the laboratory examinations of Iowa State University  (USA) \cite{pop21}and the Institute of Space and Astronautical Science (Japan) \cite{pop20} . \\

To describe the nonlinear structures, chaos theory and soliton theory are two efficient branches. Solitons consist of the most important properties of particles as well as waves, which reflect the nonlinear structures in a disciplined way. So the study of soliton solutions of nonlinear evolution equations is undeniably required. Solitons propagate with a constant configuration. Even after interaction with another soliton of the same kind, they preserved their shape, velocity, and amplitude \cite{1L1}-\cite{1L13}. Some famous nonlinear partial differential equations, like Korteweg-de Vries (KdV) equation, Kadomtsev-Petviashvili (KP) equation, and Zakharov–Kuznetsov (ZK) equation, proceed with soliton solutions. Extended and modified direct algebraic method, extended mapping method, trigonometric function method, the inverse scattering transformation \cite{11L9}, \cite{11L11} Backlund transformation, Hirota bilinear method \cite{11L14}, and Darboux transformation method \cite{11L15}   are very useful due to their efficiency in getting isolated solutions. \\
Usually solitons have elastic nature but some are non-elastic, like kink solitons, waves from Burger's equation \cite{1L14}, rational breather waves, kinky waves \cite{1L15} etc. A special type of soliton, which is a rational function solution localized in all directions of space, is known as a lump soliton \cite{1L1}- \cite{1L2}. It is a large-amplitude solitary wave, having similarities with the rough wave. Its occurrence is uncertain and vanishes without any hint. Such interesting nonlinear features are much more attractive, and the physical significance has drawn the attention of many researchers. The power of offensiveness, the devastation of sturdy nature, and the utility of signal exhibition in optical fibers, plasma physics, laser and optical physics, gas dynamics, hydrodynamics, and electromagnetics \cite{1L16} make the study of lump solitons impressive. There are many integrable nonlinear equations, like the KdV equation, the Davey-Stewartson-II equation \cite{5L1}, the three-dimensional three-wave resonant interaction equation \cite{5L2}, the BKP equation \cite{5L3}, the Kadomtsev–Petviashvili-I (KP-I) equation \cite{5L4}, those produces lump solutions \cite{2L20}-\cite{2L23}. Lump structures exist in many integrable equations \cite{12L14}-\cite{12L15}. Ma et al. \cite{ma} has derived a class of lump solutions from such an integrable equation named the KP equation using the HBM. The expansive occurrence of lump solitons in the nonlinear evolution equations, especially the KP equation, impresses the researchers \cite{epjp1}-\cite{epjp6} enough to expose their excitable nonlinear interpretation. \\\\

Using transverse perturbations while Boris Kadomtsev and Vladimir Petviashvili \cite{kp1} were investigating the one-soliton solution stability of the KdV equation, they deduced the KP equation. The KP equation is a renowned model equation in nonlinear wave theory. The nonlinearity and dispersion created by the ambiguity balanced each other in the KP equation. When the consequences of transverse direction are merged, the extended KdV equation in 2-dimensional space is converted to the KP equation. The studies \cite{kp2}-\cite{kp6} of nonlinearities in various plasma fields through the KP equation have become more interesting. Kumar and Malik \cite{4kp32}  have examined the propagation of soliton and also noted the presence of compressive soliton only per the KP equation in magnetized dusty plasma. Utilizing a quantum hydrodynamical model in an electron-ion Fermi plasma, Mushtaq et al. \cite{5kp21} investigated magneto-acoustic waves through the soliton solution of the KP equation with quantum diffraction effects. Masood et al. \cite{5kp22} analysed the low-frequency features transversely propagating magneto-acoustic waves in the KP model in condensed electron-positron-ion (e-p-i) magnetoplasma.  \\\\

The above discussions concluded that, the lump soliton solutions are well-considered in various fields in nonlinear sciences, theoretical physics, and water wave theory, yet no further investigation has been performed in plasma science. Inspired by these inquiries and fascinated by the inherent characteristics of lump structures, we have investigated the existence of lump solitons and their significance in parameter-dependent systems.

\section{Governing Equations  }
We consider the basic hydrodynamical model Eqs., as follows

\begin{equation}
\frac{\partial n_i}{\partial T}+ \nabla^\prime.(n_iv_i)=0 \label{b1}
\end{equation}
\begin{equation}
\frac{\partial v_i}{\partial T}+ (v_i.\nabla^\prime)v_i=- \frac{e\nabla^\prime \psi}{m_i}+\frac{eB_0}{m_i c}v_i \times e_z \label{b2}
\end{equation}
\begin{equation}
(\nabla^\prime)^2 \psi=-4\pi[-e n_e+e n_i- e z_d n_d] \label{b3}
\end{equation}
$n_i$ and $v_i$ are the number density and velocity of ions, $m_i$ is the mass of ions, $n_e$ and $n_d$ are the number densities of electrons and dust respectively. $\psi$ is the electrostatic  potential and $q_d = -e z_d$ is the dust charge, where $z_d$ is the dust charge number and $e$ is the elementary charge of electrons. Now  we normalized  the basic variables in Eqs.(\ref{b1})-(\ref{b3})  as follows 
\begin{equation} 
t=\Omega T, \nabla=\frac{c_s}{\Omega}\nabla^\prime, v=\frac{v_i}{c_s}, n=\frac{n_i}{n_{i0}}, \phi=\frac{e \psi}{T_e} \nonumber
\end{equation} 
Different power-law functions have been established due to the necessity of dynamics of distinct plasma situations in uniform magnetic plasma field. The power law tails are originated obviously by the effect of collision cross-section and mean free path on particle velocity and also the particle trajectories are changed amazingly. The difference between the collision cross-section and particle speed means diverges the tail of the velocity distribution from Gaussian/Maxwellian distribution. Space plasmas, plasmas in planetary magnetospheres, solar wind, magneto sheaths of magnetized planets, ionosphere, and planetary magnetospheres, \cite{b1}-\cite{b3} as well as astrophysical plasmas, are distinguished  more generally by an appropriate velocity distribution function, called q nonextensive distribution \cite{q37} and is recognized by Renyi \cite{q36}, where the entropic index q, is the degree of non extensively of the considered system. Due to the wide range covering capability of plasma situation, nonextensive distribution is rather admissible than the general Maxwellian distributio. Many researchers \cite{q38}-\cite{q45} have observed nonlinear structures including non-Maxwellian electrons distribution function. When the electrons obey non-extensive Tsallis distribution, the density of electrons is given by \\
\begin{equation}
n_e =\left[1+(q-1)\phi\right]^{\frac{(q+1)}{2(q-1)}}\label{b4}
\end{equation}
$ \implies $
\begin{equation}
n_e =1+a_1 \phi+b_1 \phi^2 +c_1 \phi^3 +\dots \label{b5}
\end{equation}
where $a_1 =\frac{(q+1)}{2}$ , $b_1 =\frac{(q+1)(3-q)}{8}$ , $c_1 =\frac{(q+1)(3-q)(5-3q)}{48}$\\
where $\lambda_e = {\left(\frac{T_e}{4\pi n_{e0} e^2}\right)}^{1/2}$ is the electron Debye length, $c_s={\left(\frac{T_e}{m_i}\right)}^{1/2}$ is the ion acoustic velocity, $\Omega=\frac{eB_0}{m_i c}$ is the ion gyrofrequency. $n_{e0}$, $n_{i0}$ are the electron and ion densities respectively in the unperturbed state.\\
\section{Normalized Equations }
Using the Eqs.(\ref{b1})-(\ref{b5}), we get the normalized Eqs. as
\begin{equation}
\frac{\partial n}{\partial t}+ \nabla.(n v)=0 \label{g1}
\end{equation}
\begin{equation}
\frac{\partial v}{\partial t}+(v.\nabla).v= -\nabla \phi +(v \times e_z)\label{g2}
\end{equation}
\begin{equation}
\nabla^2 \phi = \beta[(1+a_1 \phi +b_1 \phi^2 +c_1 \phi^3+\dots) -\delta_1 n + \delta_2]\label{g3}
\end{equation}
Where $\beta=\frac{4\pi e^2}{T_e} {r_g^2} $, $\delta_1 $ and $\delta_2  $ are the ratio of ions and dust grains to the electrons number density and $r_g=\frac{c_s}{\Omega}$ is the ion gyroradius.\\
\section{Derivation of Kadomtsev-Petviashvili (KP) Equation }
We assume that the wave is propagating in the x-z direction. Then the normalized
Eqs. reduce to
\begin{equation}
\frac{\partial n}{\partial t}+\frac{\partial(n v_x)}{\partial x}+\frac{\partial(n v_z)}{\partial z}=0 \label{kg1}
\end{equation}
\begin{equation}
\frac{\partial v_x}{\partial t}+\left( v_x \frac{\partial}{\partial x}+v_z \frac{\partial}{\partial z}\right)v_x = -\frac{\partial \phi}{\partial x}+v_y \label{kg2}
\end{equation}
\begin{equation}
\frac{\partial v_y}{\partial t}+\left(v_x \frac{\partial}{\partial x}+v_z \frac{\partial}{\partial z}\right)v_y=-v_x \label{kg3}
\end{equation}
\begin{equation}
\frac{\partial v_z}{\partial t}+\left( v_x \frac{\partial}{\partial x}+v_z \frac{\partial}{\partial z}\right)v_z =- \frac{\partial \phi}{\partial z} \label{kg4}
\end{equation}
\begin{equation}
\left(\frac{\partial^2}{\partial x^2}+\frac{\partial^2}{\partial z^2}\right) \phi=\beta[(1+a_1 \phi +b_1 \phi^2 +c_1 \phi^3 +\dots) -\delta_1 n +\delta_2] \label{kg5}
\end{equation}
To linearize the normalized  Eqs. (\ref{kg1})-(\ref{kg5}), let us write the dependent variables as the sum of equilibrium and perturbed parts, so, $ n=1+\bar{n}$, $ v_x=\bar{v_x}$, $v_y= \bar{v_y}$, $v_z=\bar{v_z}$ and $\phi=\bar{\phi}$. After linearizing, we take all perturbed variables of the form $ e^{i(k_x x+k_z z-\omega t)} $,  where $ k_x $, $ k_z $ are the wave numbers in $ x $ and $ z $ direction respectively, and $ \omega $ is the wave frequency $ (\omega << \Omega) $. This leads to \\
\begin{equation}
\begin{vmatrix}
-i\omega  &  ik_x  &  0  &  ik_z  &  0  \\
      0   &  -i\omega  &  -1  &  0  &  ik_x \\
      0  &  1  &  -i\omega  &  0  &  0  \\
      0  &  0  &  0  &  -i\omega  &  ik_z  \\
      \beta \delta_1  &  0  &  0  &  0  &  -(k^2_x + k^2_z + a_1\beta)
\end{vmatrix}
=0
\end{equation}
It gives the dispersion relation
\begin{equation}
\omega = k_z \left[ \frac{a_1}{\delta_1}+(1+ \frac{1}{\beta \delta_1})k^2_x  + (1+\frac{1}{\beta \delta_1})k^2_z \right]^{-1/2}\label{kgd}
\end{equation}
\subsection{Stretched Co-ordinates and Perturbation }
The dispersion relation guides us which stretching should be adopted for the fluid hydrodynamical model to obtain nonlinear evolution equations. On the basis of the dispersion relation, we take the stretching co-ordinates as follows
\begin{equation}
\chi= \epsilon^2 x \label{skp1}
\end{equation}
\begin{equation}
\xi =\epsilon(z-Vt) \label{skp2}
\end{equation}
\begin{equation}
\tau = \epsilon^3 t \label{skp3}
\end{equation}
where $ V $ is the phase velocity of IAW and $ \epsilon $ is a small parameter measuring the strength of the non-linearity.\\
The dependent variables are expanded as
\begin{equation}
n= 1+ \epsilon^2 n_1 +\epsilon^4 n_2 + \dots \label{sdkp1}
\end{equation}
\begin{equation}
v_x =\epsilon^3 v_{x_1}+ \epsilon^5 v_{x_2}+ \dots \label{sdkp2}
\end{equation}
\begin{equation}
v_y = \epsilon^3 v_{y_1}+ \epsilon^5 v_{y_2}+\dots \label{sdkp3}
\end{equation}
\begin{equation}
v_z = \epsilon^2 v_{z_1}+ \epsilon^4 v_{z_2}+ \dots \label{sdkp4}
\end{equation}
\begin{equation}
\phi = \epsilon^2 \phi_1 + \epsilon^4 \phi_2+ \dots \label{sdkp5}
\end{equation}
Then the normalized Eqs. (\ref{kg1})-(\ref{kg5}) can be written in terms of $ \chi $, $ \xi $ and $ \tau $ as follows
\begin{equation}
-\epsilon V \frac{\partial n}{\partial \xi}+ \epsilon^3 \frac{\partial n}{\partial \tau}+\epsilon^2 \frac{\partial(n v_x)}{\partial \chi}+ \epsilon \frac{\partial( n v_z)}{\partial \xi}=0 \label{pkp1}
\end{equation}
\begin{equation}
-\epsilon V \frac{\partial v_x}{\partial \xi}+\epsilon^3 \frac{\partial v_x}{\partial \tau} +\left(\epsilon^2 v_x \frac{\partial v_x}{\partial \chi}+ \epsilon v_z \frac{\partial v_x}{\partial \xi}\right) = -\epsilon^2 \frac{\partial \phi}{\partial \chi}+ v_y \label{pkp2}
\end{equation}
\begin{equation}
-\epsilon V \frac{\partial v_y}{\partial \xi}+\epsilon^3 \frac{\partial v_y}{\partial \tau}+ \left(\epsilon^2 v_x \frac{\partial v_y}{\partial \chi}+ \epsilon v_z \frac{\partial v_y}{\partial \xi}\right) =-v_x \label{pkp3}
\end{equation}
\begin{equation}
-\epsilon V \frac{\partial v_z}{\partial \xi}+\epsilon^3 \frac{\partial v_z}{\partial \tau}+\left(\epsilon^2 v_x \frac{\partial v_z}{\partial \chi}+ \epsilon v_z \frac{\partial v_z}{\partial \xi}\right) = -\epsilon \frac{\partial \phi}{\partial \xi} \label{pkp4}
\end{equation}
\begin{equation}
\left(\epsilon^4 \frac{\partial^2}{\partial \chi^2}+\epsilon^2 \frac{\partial^2}{\partial \xi^2}\right) \phi = \beta [(1+a_1 \phi +b_1 \phi^2 +c_1 \phi^3 +\dots) - \delta _1 n + \delta_2] \label{pkp5}
\end{equation}
Substituting (\ref{sdkp1})-(\ref{sdkp5}) into the Eqs. (\ref{pkp1})-(\ref{pkp5}) and equating the coefficients of each power of $ \epsilon $ to zero, we get the first order and second order density, velocity and potential function. After some simplifications, we obtain  \\
\begin{equation}
\frac{\partial}{\partial \xi}\left[\frac{\partial \phi_1}{\partial \tau}+A \phi_1 \frac{\partial \phi_1}{\partial \xi}+B \frac{\partial^3 \phi_1}{\partial \xi^3}\right] -C \frac{\partial^2 \phi_1}{\partial \chi^2}=0 \label{kp}
\end{equation}
where $ A= \frac{- 2b_1 V^2 +3a_1}{2a_1V} $, $ B = \frac{V}{2a_1\beta} $ and $ C=\frac{V}{2} $  and $a_1 = \frac{(1+q)}{2}$ , $b_1 =\frac{(1+q)(3-q)}{8}$\\
$V^2 = \frac{\delta_1}{a_1}$ is the phase velocity.
Eq. (\ref{kp}) is known as the KP equation.
\subsection{Standard form of the KP Equation}
We now take the alteration,  $ \xi $   by   $ \bar{\xi}A^{m_1}B^{n_1}C^{p_1} $,   $\phi_1 $   by   $ 6\bar{\phi_1}A^{m_2}B^{n_2}C^{p_2}$,  $ \chi $  by  $ \frac{1}{\sqrt{\sigma}}\bar{\chi}A^{m_3}B^{n_3}C^{p_3} $ and $\tau $ by $ \bar{\tau}$. \\
Then we get
\begin{center}
 $ m_1 =0 ,m_2 = -1, m_3 = 0 $ \\ \vspace{.1in}
$ n_1 =\frac{1}{3}, n_2 =\frac{1}{3}, n_3 =\frac{1}{6}  $ \\  \vspace{.1in}
$  p_1 =0, p_2 =0, p_3 =\frac{1}{2}  $
\end{center}
$\implies $
\begin{equation}\label{tr}
\xi = \bar{\xi}B^{\frac{1}{3}}, \phi_1=6\bar{\phi_1}A^{-1}B^{\frac{1}{3}}, \chi= \frac{1}{\sqrt{\sigma}}\bar{\chi} B^{\frac{1}{6}}C^{\frac{1}{2}}, \tau =\bar{\tau}
\end{equation}
So, the Eq. (\ref{kp}) converted to
\begin{equation*}
\frac{\partial}{\partial \bar{\xi}} \biggr[ \frac{\partial \bar{\phi_1}}{\partial \bar{\tau}}+ 6 \bar{\phi_1}\frac{\partial \bar{\phi_1}}{\partial \bar{\xi}}+ \frac{\partial^3 \bar{\phi_1}}{\partial {\bar{\xi}}^3} \biggr] - \sigma \frac{\partial^2 \bar{\phi_1}}{\partial {\bar{\chi}}^2} =0
\end{equation*}
Omitting bar and subscript, we obtain the standard KP equation as
\begin{equation}
\frac{\partial}{\partial \xi} \biggr[ \frac{\partial \phi}{\partial \tau}+ 6 \phi \frac{\partial \phi}{\partial \xi}+ \frac{\partial^3 \phi}{\partial {\xi}^3} \biggr] - \sigma \frac{\partial^2 \phi}{\partial \chi^2} =0 \label{skp}
\end{equation}
If $ \sigma =1 $, then Eq. (\ref{skp}) is called KP-I equation and when $ \sigma=-1$, Eq. (\ref{skp}) is called KP-II equation.

\section{Lump Solutions of the KP Equation }
In this section, we will derive the lump solutions for KP-I equation. \\
For this, we consider a transformation such that the dependent variable $\phi(\xi, \chi, \tau)$ converted to a new dependent variable $f(\xi, \chi,\tau)$, as follows
\begin{equation}
\begin{split}
\phi(\xi, \chi, \tau)= & 2 \frac{\partial^2}{\partial \xi^2}  \left( ln f\right) \\ & = 2 \left[ \frac{f_{\xi\xi}f-f^2_{\xi}}{f^2}\right] \label{ho1}
\end{split}
\end{equation}
Then the KP-I equation ($\sigma=1$ in (\ref{skp})), converted to the following equation
\begin{equation}
\left(f_{\xi \tau}f -f_{\xi}f_{\tau}\right) -  \left(f_{\chi\chi}f -f^2_\chi\right) +f_{\xi\xi\xi\xi}f -4f_{\xi\xi\xi}f_{\xi}+3f^2_{\xi\xi} =0 \label{ho2}
\end{equation}
Now using Hirota derivatives, we obtain the Hirota bilinear form of Eq. (\ref{ho2}) as
\begin{equation}
\left(D_{\xi} D_{\tau} + D^4_{\xi}\right)(f.f)- D^2_\chi(f.f)=0 \label{ho3}
\end{equation}
To derive lump solutions of KP-I equation, we consider the introduced dependent variable $f$ as follows
\begin{equation}\label{ho5}
f=g^2 +h^2 +d_9  ,
\qquad
g=d_1 \xi +d_2 \chi+ d_3 \tau + d_4,
\qquad
 h=d_5 \xi+d_6 \chi+d_7 \tau+d_8,
\end{equation}
\vspace{.2in} where $ d_i, 1\leq i\leq 9$, are the real parameter.\\
Using the value of $f$, we get the following set of parameters
\begin{equation}\label{ho6}
\left\{
\begin{split}
 & d_1 =d_1, d_2 =d_2, d_3 = \frac{d_1 d^2_2 -d_1 d^2_6 + 2 d_2 d_5 d_6}{d^2_1 +d^2_5}, \\ &
d_4 =d_4, d_5 =d_5, d_6 =d_6, d_7 =\frac{2d_1 d_2 d_6 -d^2_2d_5 +d_5 d^2_6}{d^2_1 +d^2_5}, \\ &
d_8 =d_8, d_9 =\frac{3(d^2_1 +d^2_5)^3}{(d_1 d_6 -_2 d_5)^2}
\end{split}
\right\}
\end{equation}
with the determinant condition
\begin{equation}\label{ho7}
\Delta:=d_1 d_6 -d_2 d_5= \begin{vmatrix}
d_1  &  d_2 \\
d_5  &  d_6
\end{vmatrix}
\neq 0
\end{equation} 
Hence the introduced dependent variable expressed as 
\begin{equation}\label{ho8}
\begin{split}
f= & \left[ d_1 \xi +d_2 \chi+ \frac{d_1 d^2_2 -d_1 d^2_6 +2d_2 d_5 d_6}{d^2_1 +d^2_5}\tau +d_4 \right]^2 \\ & + \left[d_5 \xi +d_6 \chi+\frac{2d_1 d_2 d_6 -d^2_2 d_5 +d_5 d^2_6}{d^2_1 +d^2_5}\tau +d_8 \right]^2 \\ & + \frac{3(d^2_1 +d^2_5)^3}{(d_1 d_6 -d_2 d_5)^2},
\end{split}
\end{equation}
and the lump solutions of KP-I equation are given by 
\begin{equation}
\phi(\xi,\chi,\tau)=\frac{4(d^2_1 +d^2_5)f-8(d_1 g +d_5 h)^2}{f^2}
\end{equation}
where the expression of $f$ is given in Eq. (\ref{ho8}), and $g$ and $h$ are as follows
\begin{equation}
g=d_1 \xi +d_2 \chi +\frac{d_1 d^2_2-d_1 d^2_6 +2d_2  d_5 d_6}{d^2_1 +d^2_5}\tau +d_4,
\end{equation}
\begin{equation}
h=d_5 \xi +d_6 \chi+\frac{2d_1 d_2 d_6-d^2_2 d_5 +d_5 d^2_6}{d^2_1 +d^2_5}\tau +d_8.
\end{equation}
Now the lump solutions of the Eq. (\ref{kp}), as follows
\begin{equation}
\phi(\xi,\chi,\tau)= 4 (\frac{6B^{1/3}}{A})  \left[\frac{(d_1^2 +d_5^2)f -2(d_1 g +d_5 h)^2}{f^2}\right] \label{lu1}
\end{equation}
where  $g$ , $h$ and $f$ are given by
\begin{equation}
g=\frac{1}{B^{1/3}C^{1/2}}\left[C^{1/2}d_1 \xi + B^{1/6}d_2 \chi+B^{1/3}C^{1/2} d_3 \tau + B^{1/3}C^{1/2}d_4 \right] \label{lu2}
\end{equation}
\begin{equation}
h= \frac{1}{B^{1/3}C^{1/2}}\left[C^{1/2}d_5 \xi +B^{1/6}d_6 \chi + B^{1/3}C^{1/2}d_7 \tau +B^{1/3}C^{1/2}d_8\right] \label{lu3}
\end{equation}

and
\begin{equation}\label{lu4}
\begin{split}
f= &  \frac{1}{B^{2/3}C} \Bigg( \left( C^{1/2}d_1 \xi +B^{1/6}d_2 \chi + B^{1/3}C^{1/2}d_3 \tau +B^{1/3}C^{1/2}d_4 \right)^2 \\ & + \left( C^{1/2}d_5 \xi + B^{1/6}d_6 \chi + B^{1/3}C^{1/2}d_7 \tau + B^{1/3}C^{1/2}d_8 \right)^2 \\ & + B^{2/3}C d_9 \Bigg)
\end{split}
\end{equation}
where
\[
A=\frac{-2b_1 V^2 + 3a_1}{2a_1 V}, B= \frac{V}{2a_1 \beta}, C=\frac{-V}{2},
\]
\[
a_1 = \frac{(1+q)}{2}, b_1 = \frac{(1+q)(3-q)}{8}
\]
\section*{Particular Parameter Value  }
\textbf{Set 1:}  $d_1 =1, d_2 =a, d_5 =0, d_6 =b, d_4 =c, d_8 =d $.
Then
\begin{equation}\label{s1lu1}
g=\frac{\xi}{B^{1/3}}+ \frac{a\chi}{B^{1/6}C^{1/2}}+ (a^2 -b^2)\tau +c
\end{equation}
\begin{equation}
h=\frac{b\chi}{B^{1/6}C^{1/2}}+2ab \tau + d \label{s1lu2}
\end{equation}
\begin{equation}\label{s1lu3}
\begin{split}
f=  &  \left( \frac{\xi}{B^{1/3}}+ \frac{a\chi}{B^{1/6}C^{1/2}}+(a^2 -b^2)\tau + c \right)^2 \\ & + \left(\frac{b\chi}{B^{1/6}C^{1/2}}+ 2ab \tau + d \right)^2 + \frac{3}{b^2}
\end{split}
\end{equation}
Hence the lump solutions 
\begin{equation} \label{s1lu5}
\phi(\xi, \chi, \tau)= 4 (\frac{6B^{1/3}}{A}) \frac{ -\left(\frac{\xi}{B^{1/3}} + \frac{a \chi}{B^{1/6}C^{1/2}}+ (a^2 -b^2)\tau +c \right)^2 + \left( \frac{b \chi}{B^{1/6}C^{1/2}} +2ab \tau + d \right)^2 + \frac{3}{b^2}}  { \Bigg( \left( \frac{\xi }{B^{1/3}}+ \frac{a\chi}{B^{1/6}C^{1/2}}+ (a^2 -b^2) \tau + c\right)^2 + \left( \frac{b\chi}{B^{1/6}C^{1/2}}+ 2ab \tau + d \right)^2 + \frac{3}{b^2} \Bigg) ^2}
\end{equation}
Under the consideration of $c=d=0$, (\ref{s1lu5}) reduces to 
\begin{equation}\label{s1lump}
\phi(\xi, \chi ,\tau)= 4 (\frac{6B^{1/3}}{A}) \frac{-\left( \frac{\xi}{B^{1/3}}+ \frac{a\chi}{B^{1/6}C^{1/2}}+ (a^2 -b^2)\tau \right)^2 + b^2 \left( \frac{\chi}{B^{1/6}C^{1/2}}+ 2a\tau \right)^2 + \frac{3}{b^2}}{ \Bigg( \left( \frac{\xi}{B^{1/3}}+ \frac{a\chi}{B^{1/6}C^{1/2}}+ (a^2 -b^2)\tau \right)^2 + b^2 \left( \frac{\chi}{B^{1/6}C^{1/2}}+ 2a \tau \right)^2+ \frac{3}{b^2} \Bigg)^2 }
\end{equation}
\textbf{Set 2:}  $d_1 =1, d_2 =2, d_4 =0, d_5= 1, d_6 =-1, d_8 =0 $. Then
\begin{equation}\label{s2lu1}
g=\frac{\xi}{B^{1/3}}+ \frac{2\chi}{B^{1/6}C^{1/2}} - \frac{1}{2} \tau
\end{equation}
\begin{equation}\label{s2lu2}
h= \frac{\xi}{B^{1/3}}- \frac{\chi}{B^{1/6}C^{1/2}}- \frac{7}{2}\tau
\end{equation}
\begin{equation}\label{s2lu3}
f= \frac{25}{2}\tau^2 -8 \frac{\xi \tau}{B^{1/3}} + 5 \frac{\chi \tau}{B^{1/6}C^{1/2}}+ 2 \frac{\xi^2}{B^{2/3}}+ 5 \frac{\chi^2}{B^{1/3}C}+ 2 \frac{\xi \chi}{B^{1/2}C^{1/2}}+ \frac{8}{3}
\end{equation}
So, the lump solutions 
\begin{equation}\label{s2lump}
\phi(\xi, \chi ,\tau)= 48(\frac{6B^{1/3}}{A}) \frac{\left( -21 \tau^2 + 48 \frac{\xi \tau}{B^{1/3}} + 78 \frac{\chi \tau}{B^{1/6}C^{1/2}} -12 \frac{\xi^2}{B^{2/3}}+ 24\frac{\chi^2}{B^{1/3}C}- 12 \frac{\xi \chi}{B^{1/2}C^{1/2}}+ 16 \right)}{ \left( 75 \tau^2 -48 \frac{\xi \tau}{B^{1/3}}+ 30 \frac{\chi \tau }{B^{1/6}C^{1/2}}+ 12 \frac{\xi^2}{B^{2/3}}+ 30 \frac{\chi^2}{B^{1/3}C} + 12 \frac{\xi \chi}{B^{1/2}C^{1/2}}+ 16 \right)^2}
\end{equation}
\textbf{Set 3:} $d_1 = 1, d_2 =-2, d_4 =0, d_5 =-2, d_6 =1, d_8 =0. $ Then
\begin{equation}\label{s3lu1}
g= \frac{\xi}{B^{1/3}} -\frac{2\chi}{B^{1/6}C^{1/2}}+ \frac{11}{5} \tau
\end{equation}
\begin{equation}\label{s3lu2}
h= \frac{-2\xi}{B^{1/3}}+ \frac{\chi}{B^{1/6}C^{1/2}}+ \frac{2}{5}\tau
\end{equation}
\begin{equation}\label{s3lu3}
f= \frac{5\xi^2}{B^{2/3}}+ \frac{5 \chi^2}{B^{1/3}C}+ 5\tau^2 - \frac{8 \xi \chi }{B^{1/2}C^{1/2}} -\frac{8 \chi \tau}{B^{1/6}C^{1/2}} + \frac{14}{5} \frac{\xi \tau}{B^{1/3}} + \frac{125}{3}
\end{equation}
Hence the lump solutions are given as follows
\begin{equation}\label{s3lump}
\phi(\xi, \chi ,\tau)= 12(\frac{6B^{1/3}}{A}) \frac{1581 \tau^2 - \frac{1875\xi^2}{B^{2/3}} - \frac{525 \chi^2}{B^{1/3}C}+ \frac{3000 \xi \chi}{B^{1/2}C^{1/2}} -  \frac{1320 \chi \tau}{B^{1/6}C^{1/2}} - \frac{1050 \xi \tau}{B^{1/3}} + 15625}{\left(     \frac{75 \xi^2}{B^{2/3}}+ \frac{75 \chi^2}{B^{1/3}C} + 75 \tau^2 - \frac{120 \xi \chi }{B^{1/2}C^{1/2}}- \frac{120 \chi \tau }{B^{1/6}C^{1/2}} + \frac{42 \xi \tau }{B^{1/3}} + 625 \right)^2}
\end{equation}
\section{Results and Discussions }
In this work, a renowned nonlinear evolution equation, known as KP equation, has been derived by applying Reductive Perturbation Technique (RPT) from a dust hydrodynamical model. Then using the HBM, lump soliton solutions have been derived. Here three different set of parameter values have been chosen and accordingly, three lump soliton solutions arrived. Next we plot these solutions and  discuss the effects of plasma parameters on these.

\begin{figure}[H]
	\centering
	\subfigure[]{\includegraphics[width=0.3\linewidth]{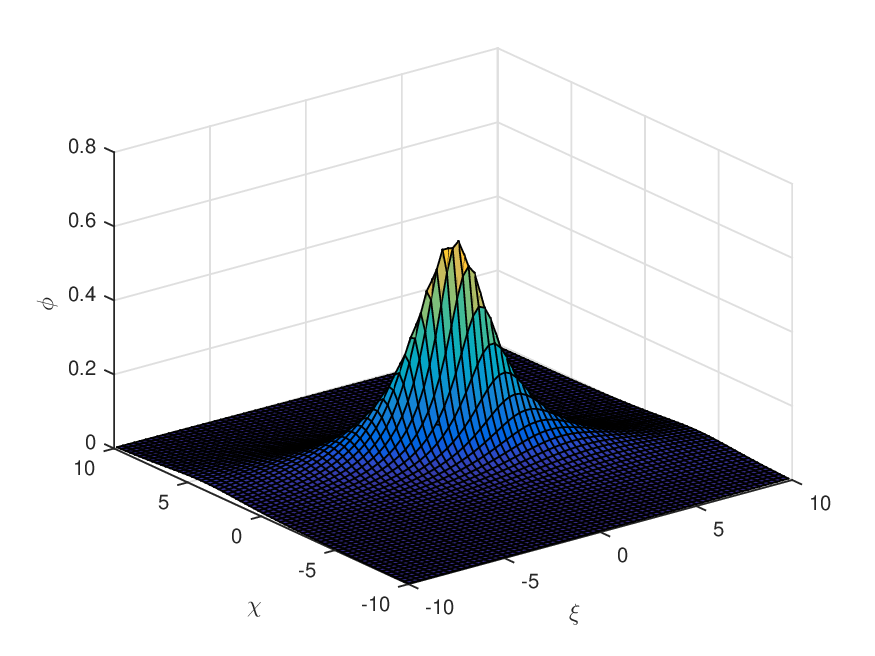}\label{f:5b}}
	\subfigure[]{\includegraphics[width=0.3\linewidth]{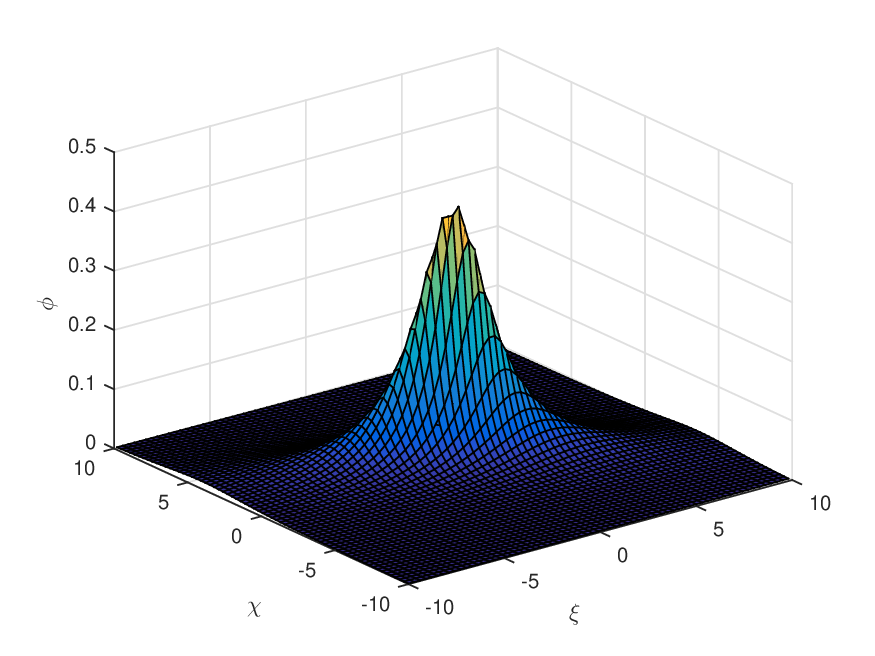}\label{f:5d}}
    \subfigure[]{\includegraphics[width=0.3\linewidth]{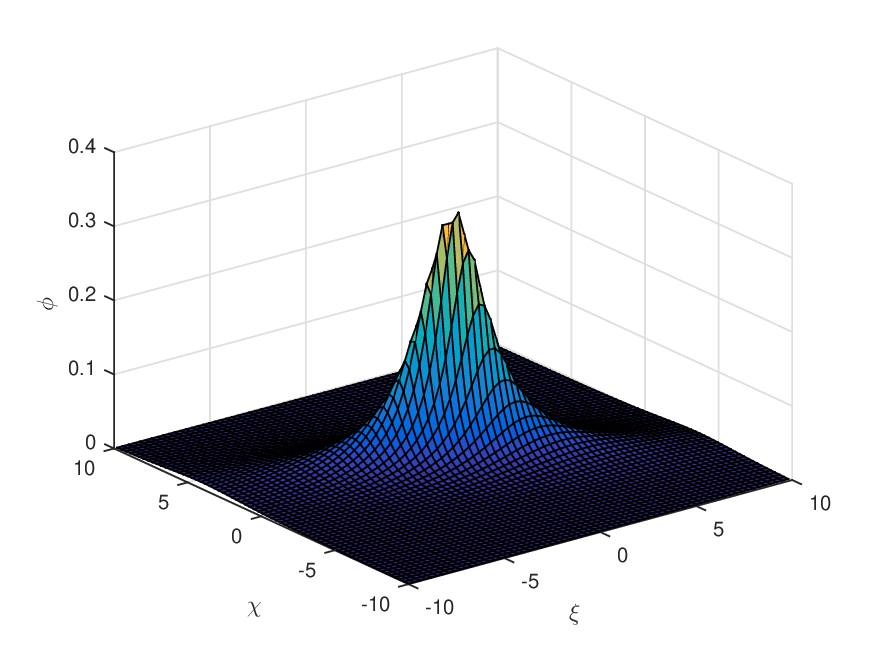}\label{f:5d}}
	\caption{Lump soliton solutions for the parameters, $a=1$, $b=1$, $\beta =0.2$, $\delta_1=1.005$, $q=1.05$, $q=1.45$, $q=1.85$ at $\tau=0$ }
\end{figure}
\begin{figure}[H]
	\centering
	\subfigure[]{\includegraphics[width=0.3\linewidth]{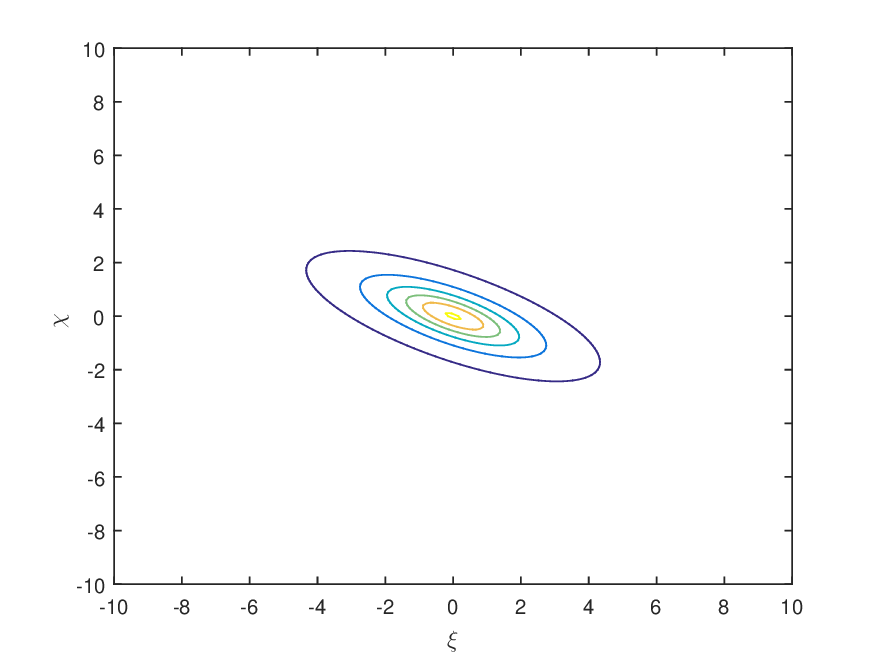}\label{f:5b}}
	\subfigure[]{\includegraphics[width=0.3\linewidth]{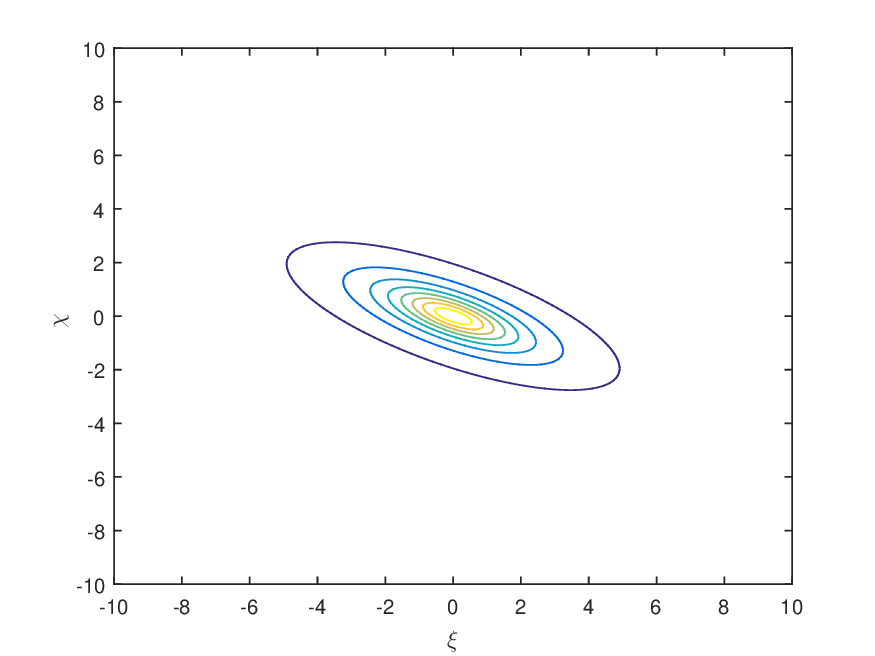}\label{f:5d}}
    \subfigure[]{\includegraphics[width=0.3\linewidth]{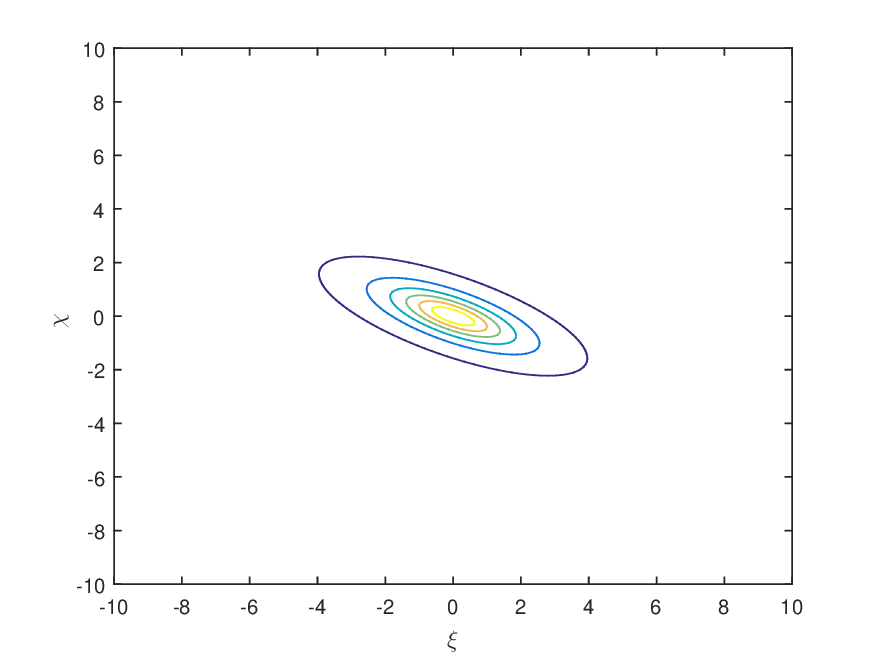}\label{f:5d}}
	\caption{Contour plot of lump solutions, for $a=1$, $b=1$,$\beta =0.2$, $\delta_1=1.005$, $q=1.05$, $q=1.45$, $q=1.85$ at $\tau=0$}
\end{figure}
\begin{figure}[H]
	\centering
	\subfigure[]{\includegraphics[width=0.3\linewidth]{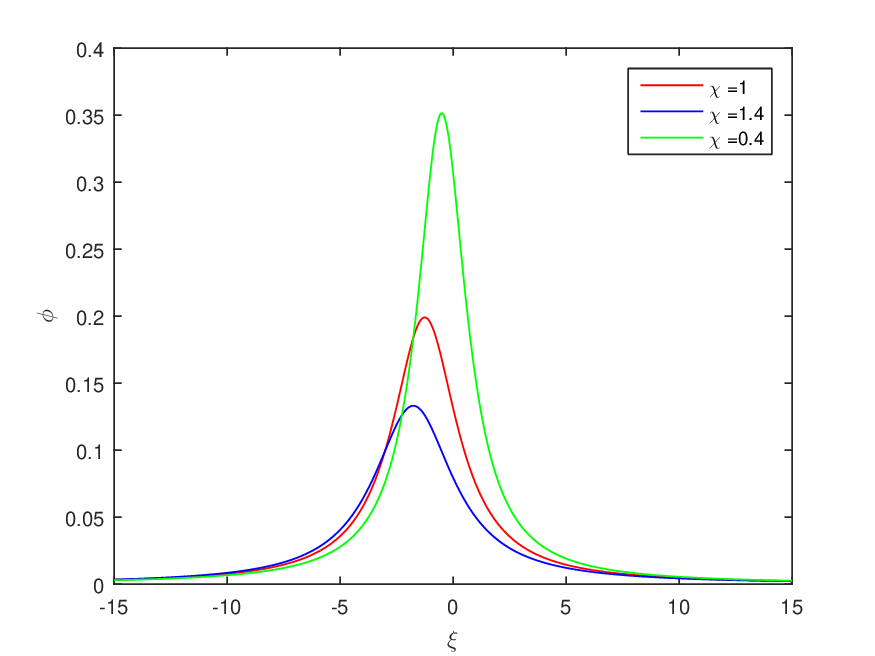}\label{f:5b}}
	\subfigure[]{\includegraphics[width=0.3\linewidth]{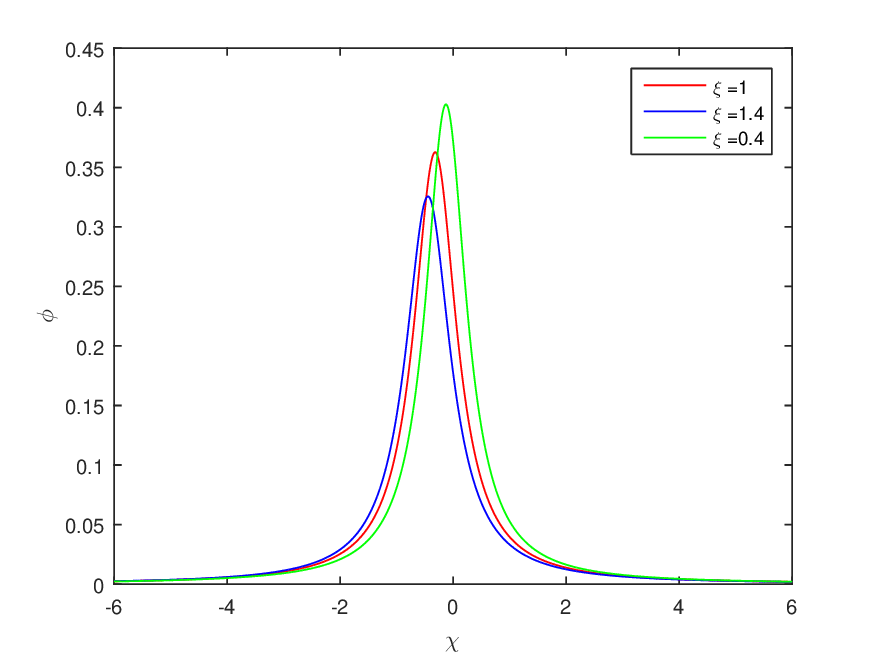}\label{f:5d}}
    \caption{3(a) and 3(b) are the 2-D plot of lump soliton solutions for $\phi $ vs $\xi $, and  $\phi $ vs $\chi $ respectively, for $\beta =0.2$, $\delta_1=1.005$, $q=1.55$ at $\tau=0$ }
\end{figure}

For parameter set-1, lump solitons have been plotted for distinct values of nonextensive parameter q at time $\tau=0$, remaining the other parameters as fixed. Figures 1(a), 1(b), and 1(c) are the 3-D lump structures, and  Figure 2(a), 2(b), and  2(c) describe the solution features in the phase diagram for same parameter set.    Figure (1) narrates that the amplitude of lump solitons are gradually decreasing for increasing value of nonextensive parameter $q$. Figure (2) describes the region where the system is close as well as conservative for same parameter regime. Figures 3(a) and 3(b) are plotted to explain the behavioural change of lump solitons more conveniently, for $q=1.55$.Figure 3(a) is framed by $\phi$ and $\xi$ co-ordinates, for $\chi =0.4$ (green line), $\chi =1$ (red line), $\chi =1.4$ (blue line).  Similarly, Figure 3(b) is plotted for $\phi$ with respect to the coordinate axis $\chi $ for $\xi =0.4$ (green line), $\xi =1$ (red line), $\xi =1.4$ (blue line).

\begin{figure}[H]
	\centering
	\subfigure[]{\includegraphics[width=0.3\linewidth]{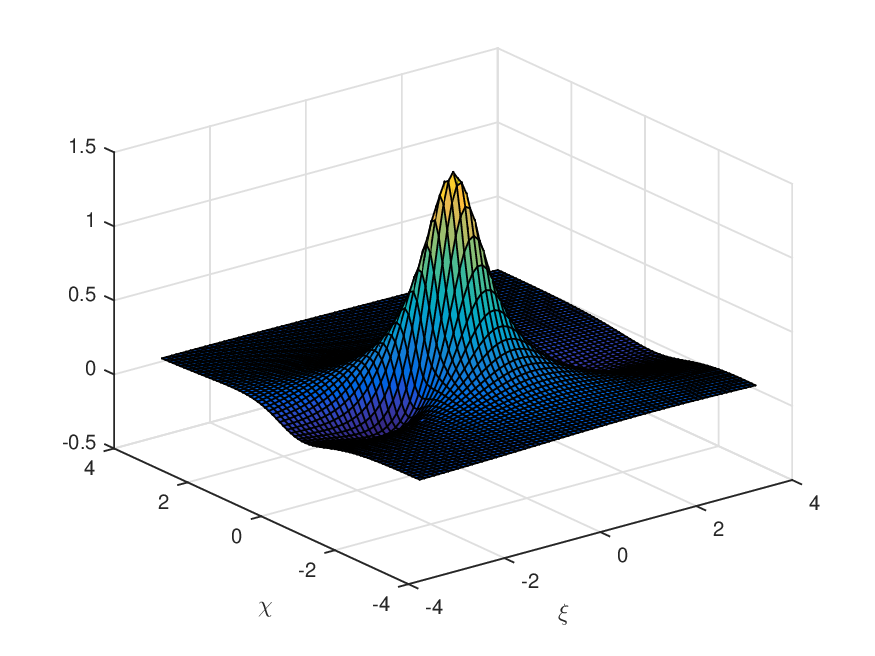}\label{f:5b}}
	\subfigure[]{\includegraphics[width=0.3\linewidth]{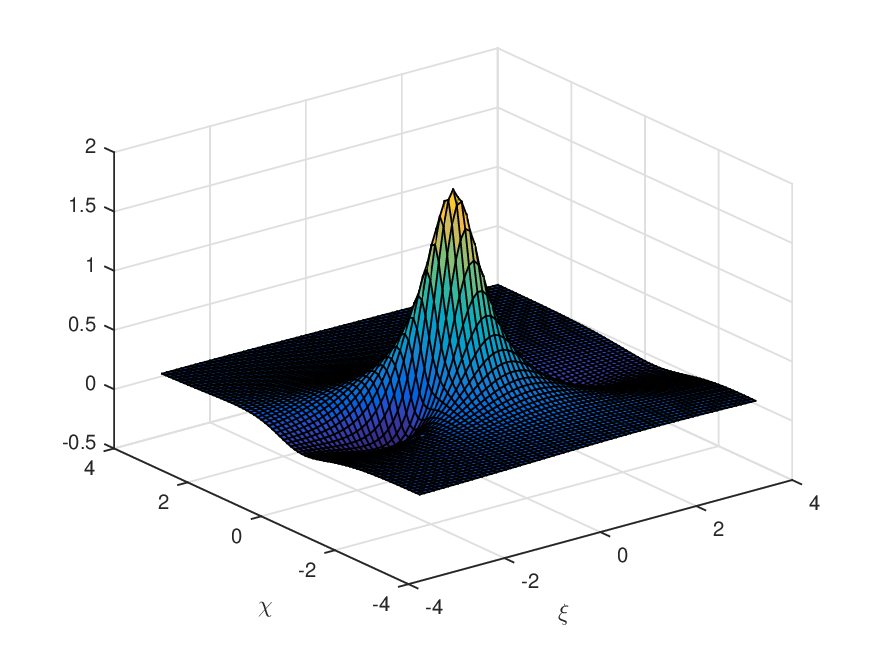}\label{f:5d}}
    \subfigure[]{\includegraphics[width=0.3\linewidth]{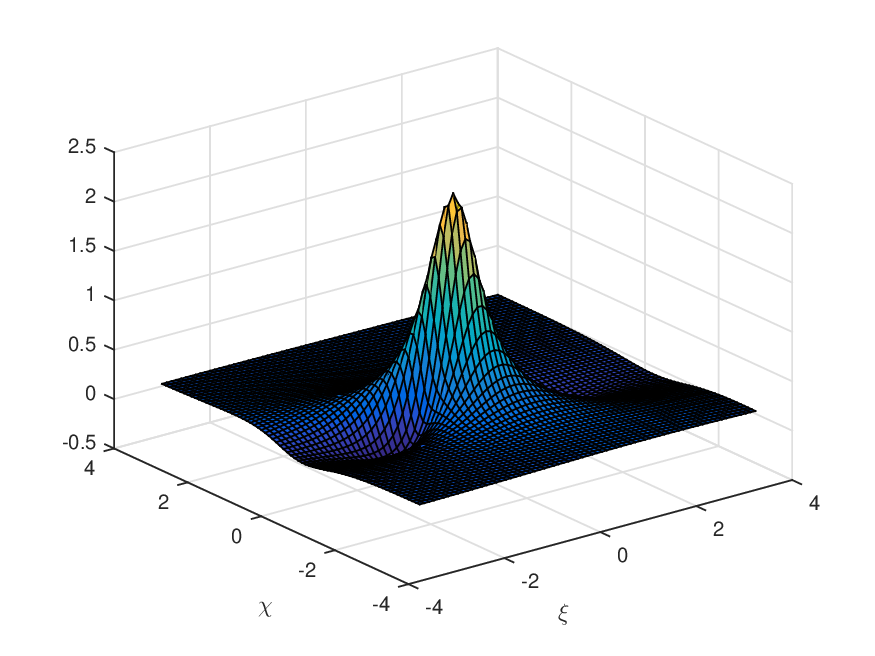}\label{f:5d}}
	\caption{Lump soliton solutions, for the parameters, $\beta =0.2$, $\delta_1=1.005$, $q=1.05$, $q=1.45$, $q=1.85$ at $\tau=0$}
\end{figure}
\begin{figure}[H]
	\centering
	\subfigure[]{\includegraphics[width=0.3\linewidth]{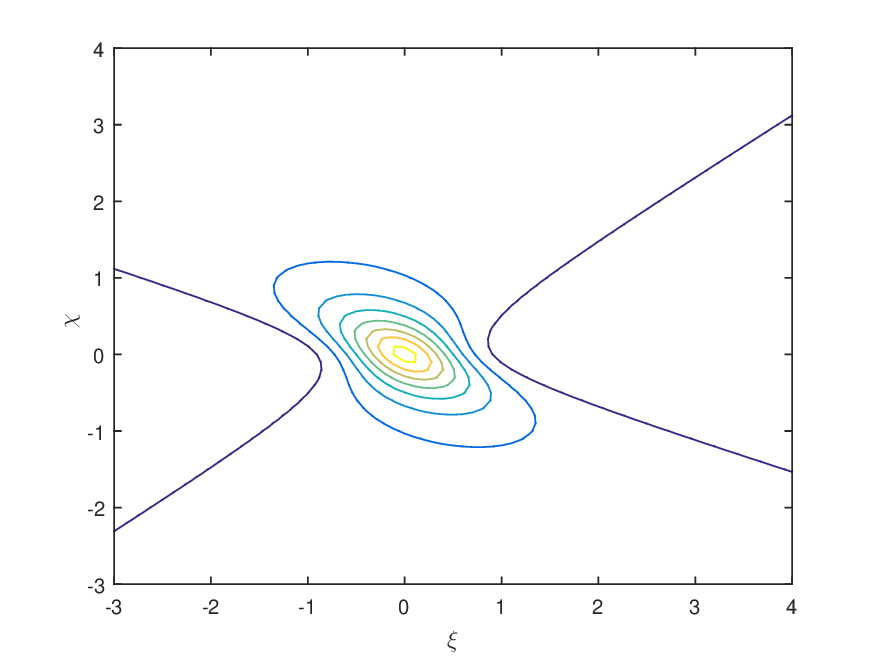}\label{f:5b}}
	\subfigure[]{\includegraphics[width=0.3\linewidth]{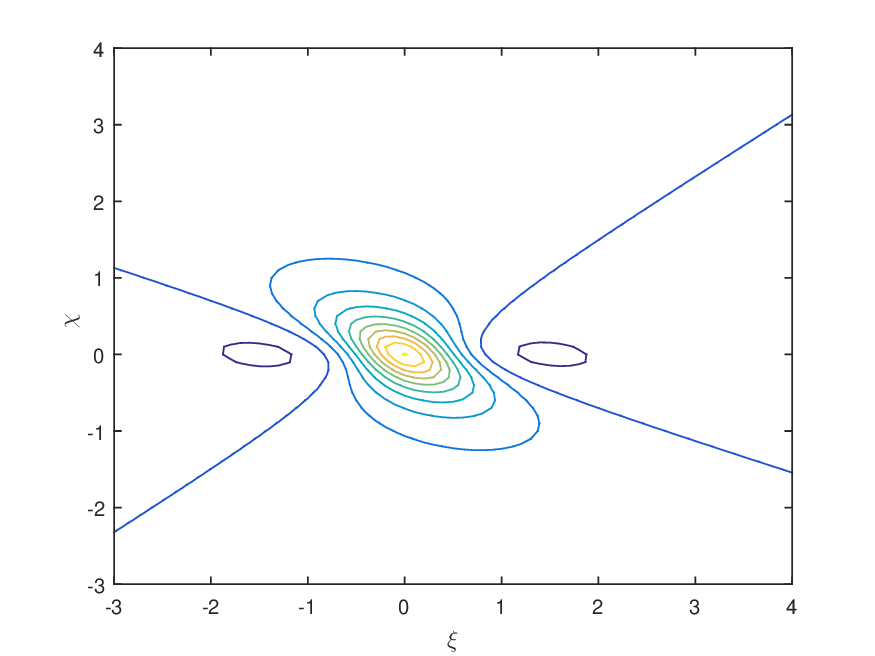}\label{f:5d}}
    \subfigure[]{\includegraphics[width=0.3\linewidth]{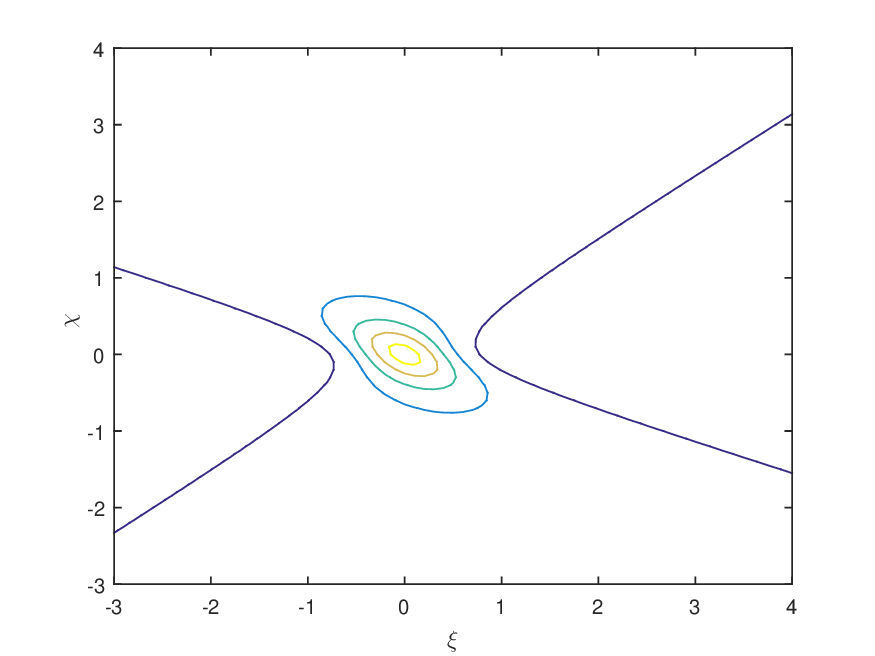}\label{f:5d}}
	\caption{Contour plot of lump solitons, for $\beta =0.2$, $\delta_1=1.005$, $q=1.05$, $q=1.45$, $q=1.85$ at $\tau=0$}
\end{figure}
\begin{figure}[H]
	\centering
	\subfigure[]{\includegraphics[width=0.3\linewidth]{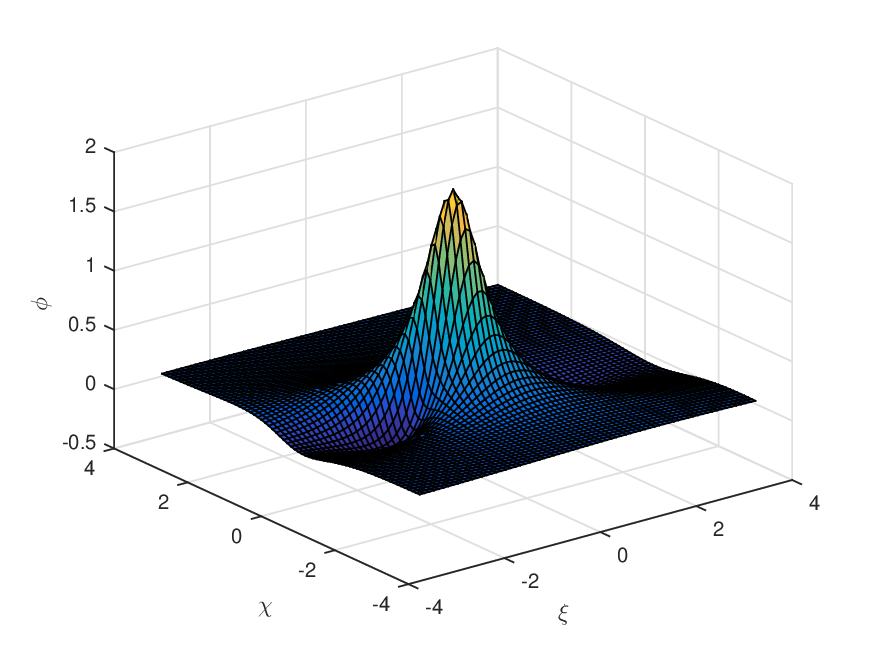}\label{f:5b}}
	\subfigure[]{\includegraphics[width=0.3\linewidth]{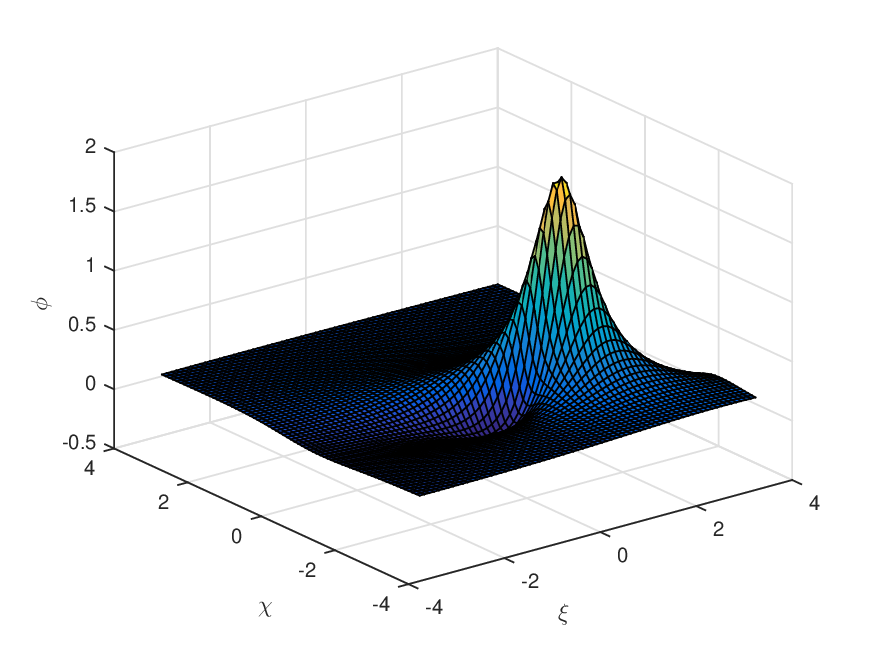}\label{f:5d}}
    \subfigure[]{\includegraphics[width=0.3\linewidth]{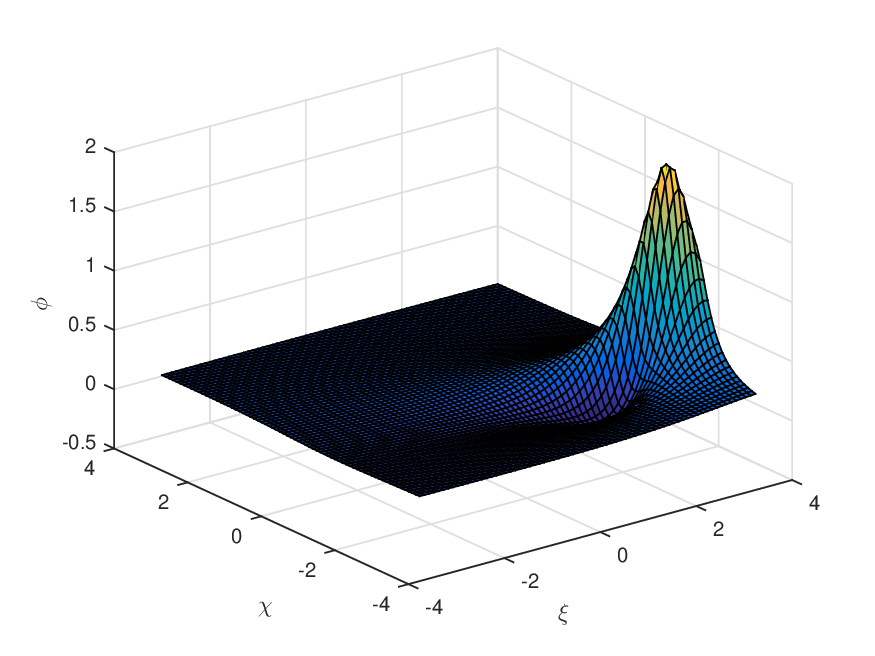}\label{f:5d}}
	\caption{Lump soliton solutions, for the parameters, $\beta =0.2$, $\delta_1=1.005$, $q=1.85$ at $\tau=0$, $\tau=1$, $\tau=2$}
\end{figure}
\begin{figure}[H]
	\centering
	\subfigure[]{\includegraphics[width=0.3\linewidth]{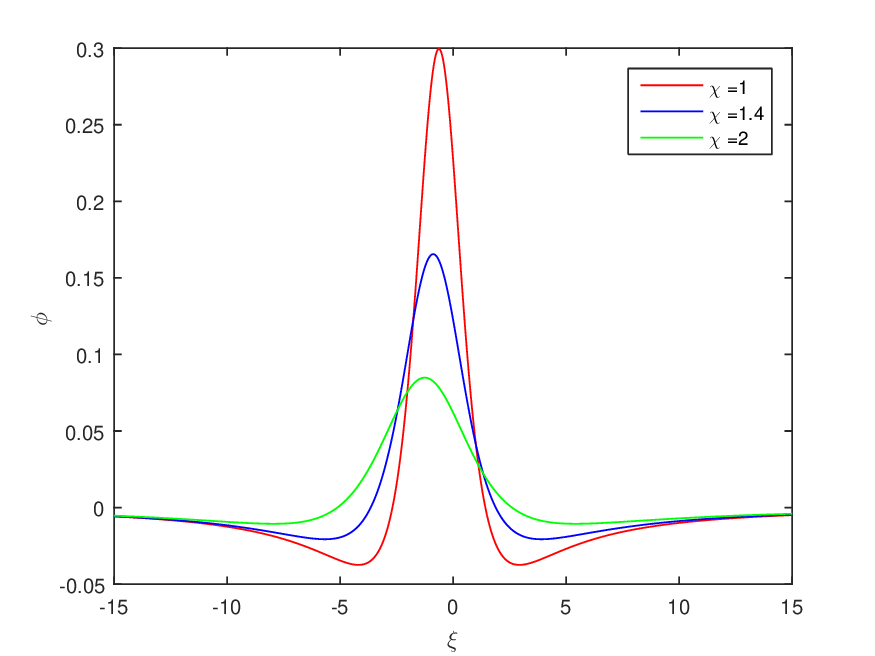}\label{f:5b}}
	\subfigure[]{\includegraphics[width=0.3\linewidth]{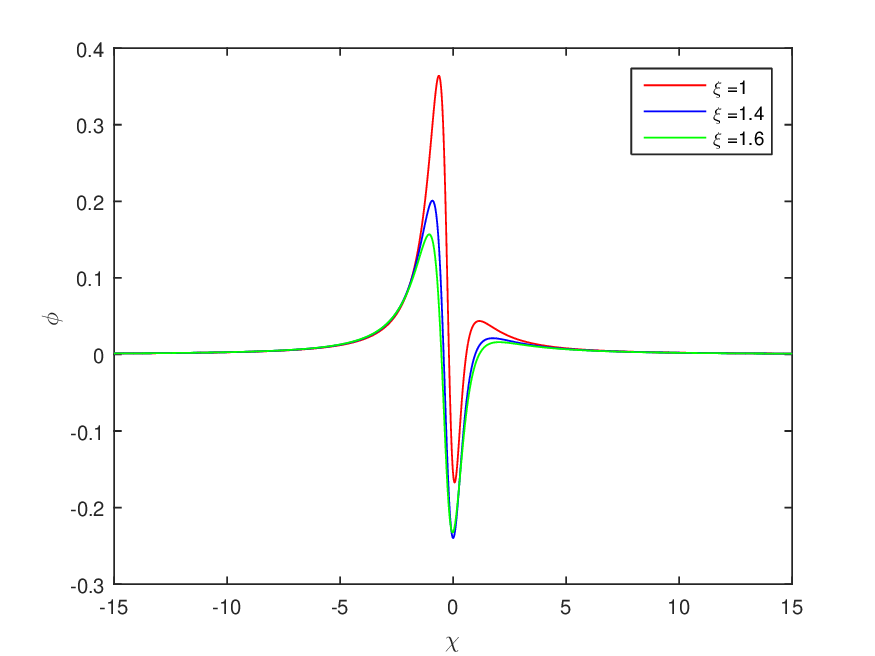}\label{f:5d}}
    \caption{7(a) and 7(b) are the 2-D profile of lump solitons, for $\phi $ vs $\xi $ and  $\phi $ vs $\chi $ respectively, for $\beta =0.2$, $\delta_1=1.005$, $q=1.55$ at $\tau=0$}
\end{figure}
Figures (4)-(7) have plotted for parameter set-2, to see the features of lump solitons for distinct values of nonextensive parameter q, keeping rest of the parameter as constant at $\tau=0$. Figure (4) describes the lump solitons and Figure (5)  shows the solution features in the phase diagram for $q=1.05$, $q=1.45$, $q=1.85$ respectively. From Figure (4), it has been observed that, the amplitude of lump solitons are gradually increasing for increasing values of nonextensive parameter $q$. Figure (5) displays the region where the system is close as well as conservative for same parameter regime. Figure (6) is the lump structures at various time $\tau=0$, $\tau=1$, and $\tau=2$ respectively. Figures 7(a), and 7(b) are the 2-D profile of lump structures, describe strongly the behaviourial change of solitons for $q=1.55$. In Figure 7(a), the electrostatic potential ($\phi$) has been drawn with respect to the co-ordinate  axis $\xi$, for $\chi =2$ (green line), $\chi =1$ (red line), $\chi =1.4$ (blue line).  Similarly, we have plotted $\phi$ vs $\chi$ graph in 7(b), for different frame i.e. for $\xi =1.6$ (green line), $\xi =1$ (red line), $\xi =1.4$ (blue line).

\begin{figure}[H]
	\centering
	\subfigure[]{\includegraphics[width=0.3\linewidth]{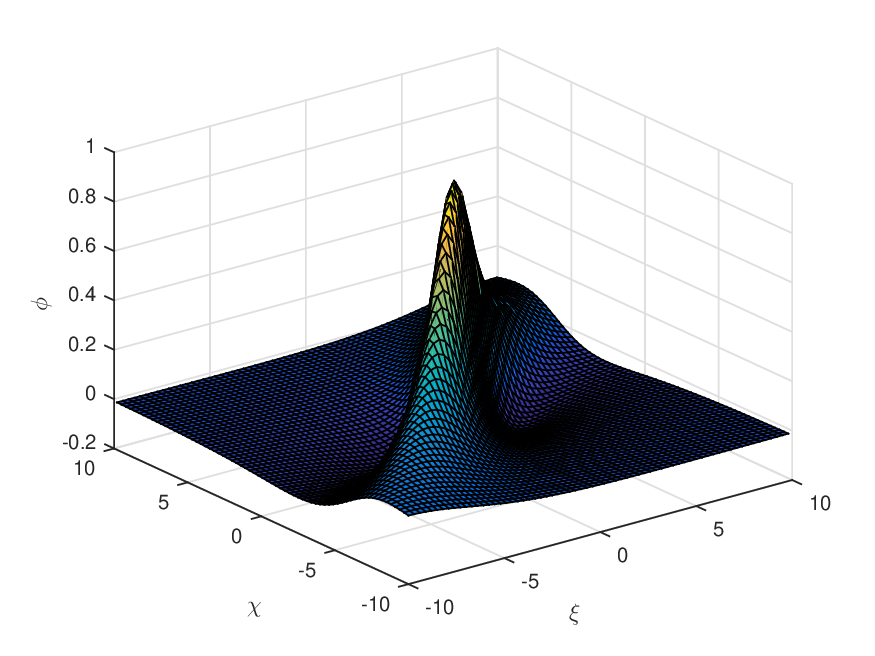}\label{f:5b}}
	\subfigure[]{\includegraphics[width=0.3\linewidth]{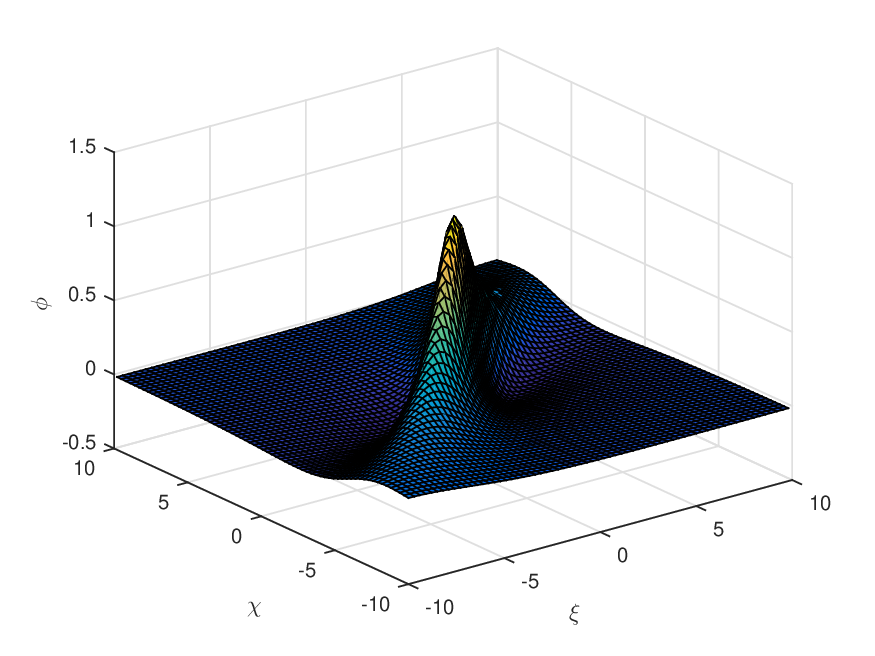}\label{f:5d}}
\caption{Lump soliton solutions, for the parameters, $\beta =0.2$, $\delta_1=1.005$, $q=1.05$, $q=1.45$, at $\tau=0$}
\end{figure}
\begin{figure}[H]
	\centering
	\subfigure[]{\includegraphics[width=0.3\linewidth]{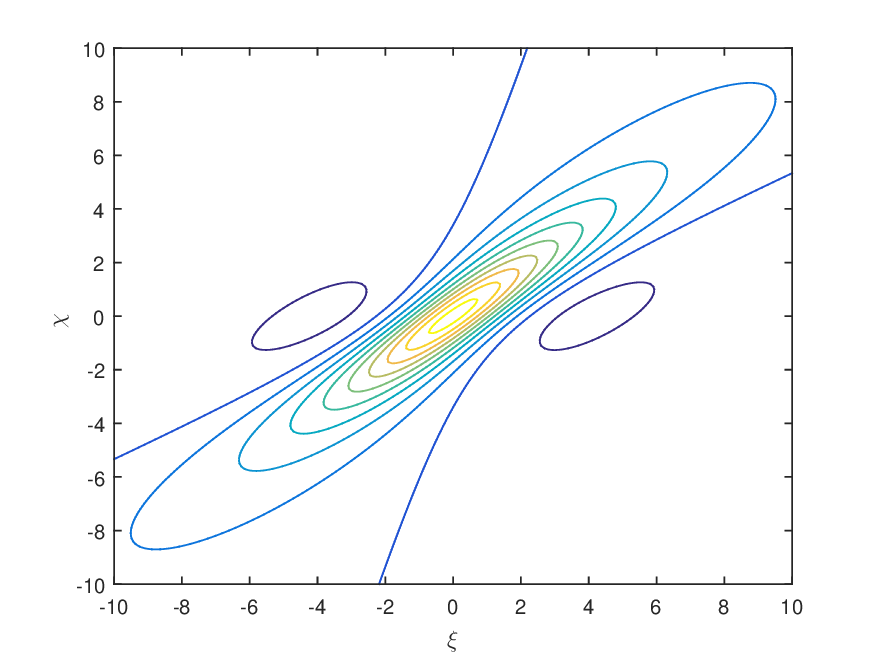}\label{f:5b}}
	\subfigure[]{\includegraphics[width=0.3\linewidth]{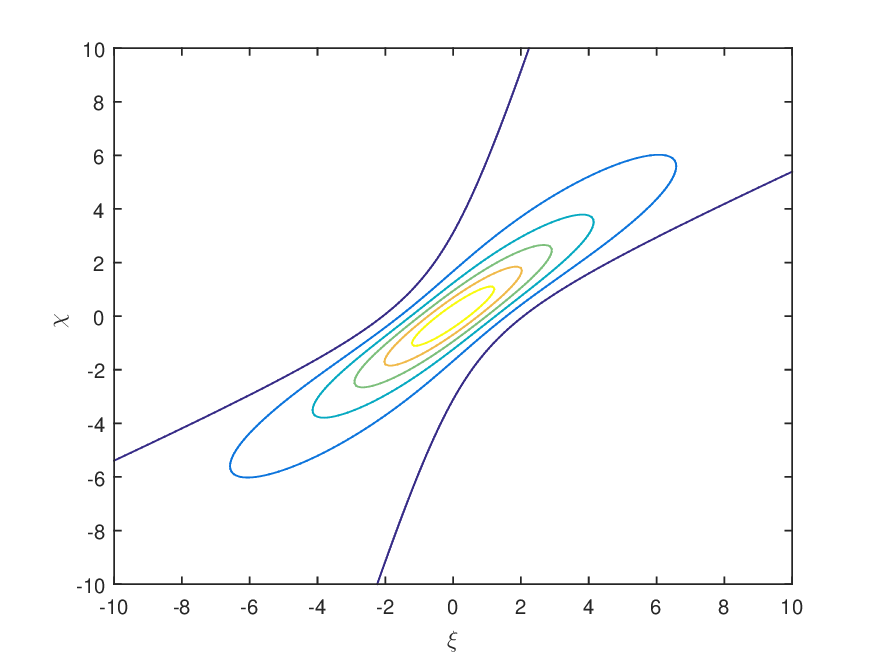}\label{f:5d}}
\caption{Contour plot of lump solitons, for $\beta =0.2$, $\delta_1=1.005$, $q=1.05$, $q=1.45$, at $\tau=0$}
\end{figure}
\begin{figure}[H]
	\centering
	\subfigure[]{\includegraphics[width=0.3\linewidth]{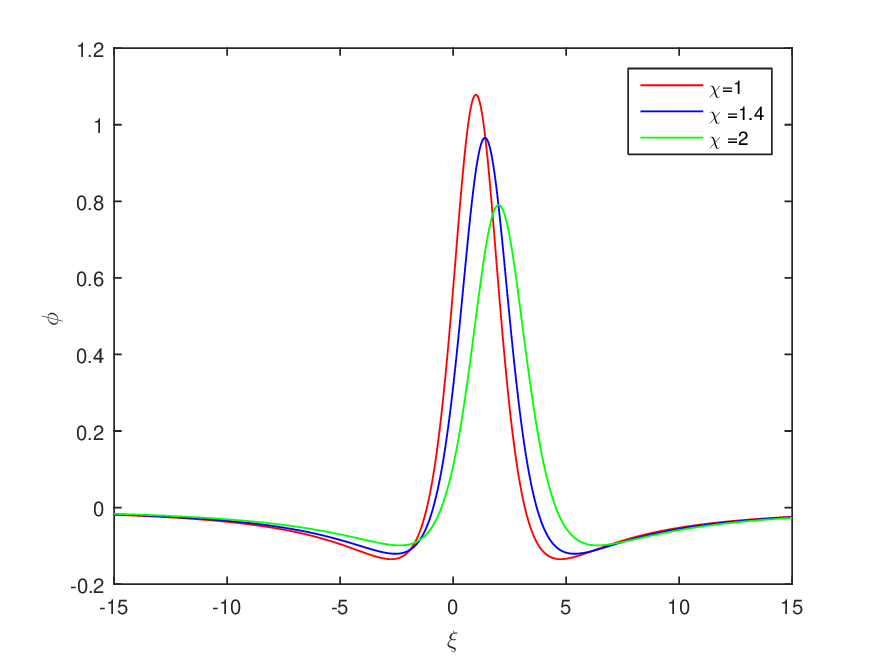}\label{f:5b}}
	\subfigure[]{\includegraphics[width=0.3\linewidth]{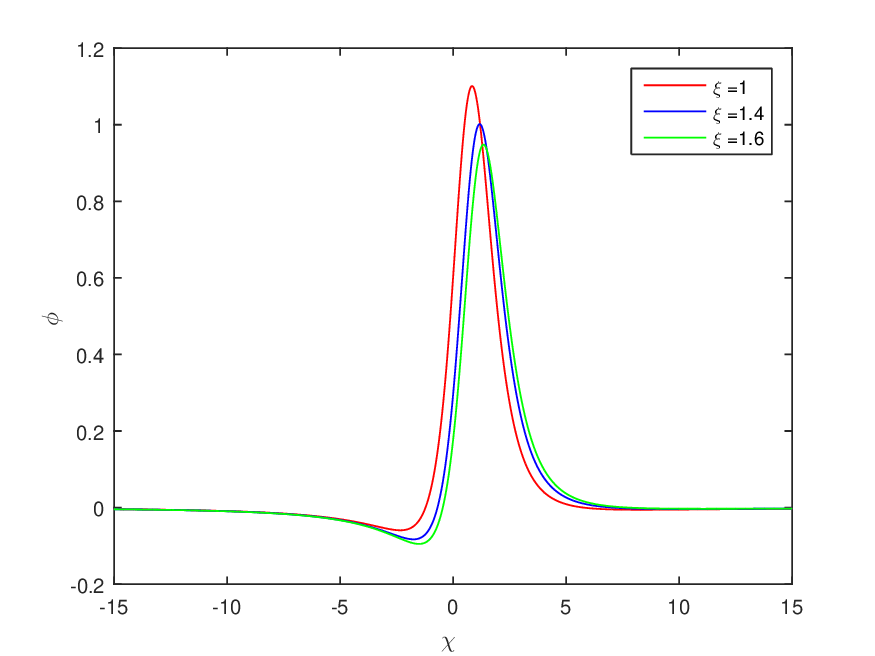}\label{f:5d}}
    \caption{10(a) and 10(b) are the 2-D profile of lump solitons, for $\phi $ vs $\xi $ and  $\phi $ vs $\chi $ respectively,  for $\beta =0.2$, $\delta_1=1.005$, $q=1.55$ at $\tau=0$}
\end{figure}

For parameter set-3, we have plotted Figures (8)-(10), to see the lump solitons for various  non-extensive \\ parameter q at time $\tau=0$ and  keeping rest of the parameters as constant. Figure 8(a), and 8(b) are the lump soliton structures and  Figure 9(a), and 9(b) are the phase diagram for $q=1.05$, and $q=1.45$ respectively. It is clear from Figures 8(a), and 8(b) that, when the nonextensive  parameter increases, the amplitude of lump solitons decreases gradually. Here Figures 9(a), and 9(b) show the region where the system is close as well as conservative for same parameter set. To study the soliton structures in more convenient way, we  have plotted the soliton structures in one space dimension. To characterize the change of soliton structures effectively, Figure (10) has been plotted, for $q=1.55$ in one space dimension only. Figure 10(a) is the $\phi$ vs $\xi$ plot, for $\chi =2$ (green line), $\chi =1$ (red line), $\chi =1.4$ (blue line) and Figure 10(b) is  the $\phi$ vs $\chi$ graph, for $\xi =1.6$ (green line), $\xi =1$ (red line), $\xi =1.4$ (blue line).\\

\Large {\textbf{DATA AVAILABILITY}}\\\\
Data sharing is not applicable to this article as no new data were created or analyzed in this study.

\end{document}